\documentclass[%
reprint, floatfix,
 amsmath,amssymb,
 aps,
 prl,
]{revtex4-1}

\usepackage{graphicx}
\usepackage{dcolumn}
\usepackage{bm}
\usepackage{hyperref}

\usepackage{xcolor}

\begin{document}

\title{Global Optimization, Local Adaptation, and the Role of Growth in Distribution Networks}

\author{Henrik Ronellenfitsch}
\email{henrikr@sas.upenn.edu}
\author{Eleni Katifori}%
 \email{katifori@sas.upenn.edu}
\affiliation{%
Department of Physics and Astronomy, University of Pennsylvania, Philadelphia, PA 19104, USA.}%

\date{\today}

\begin{abstract}
Highly-optimized complex transport networks serve crucial functions in many man-made and 
natural systems such as power grids and plant or animal vasculature. Often,
the relevant optimization functional is non-convex and characterized by many local extrema.
In general, finding the global, or nearly global optimum is difficult. 
In biological systems, it is believed that such an optimal state is slowly achieved
through natural selection.
However, general coarse grained models for flow networks with local positive feedback rules for the vessel conductivity typically get trapped in low efficiency, local minima. In this work we show how the growth of the underlying tissue, coupled to the dynamical equations for network development, can drive the system to a dramatically improved optimal state. This general model provides a surprisingly simple explanation for the appearance of highly optimized transport networks in biology such as leaf and animal vasculature.
\end{abstract}

\maketitle



Complex life requires distribution networks: veins and arteries in animals, xylem and phloem in plants, and even fungal mycelia that deliver nutrients and collect the by-products of metabolism. Efficient function of these distribution networks is crucial for an organism's fitness. Thus, biological transport networks are thought to have undergone a process of gradual optimization through evolution~\cite{DeVisser2014}, culminating in organizational
principles such as Murray's law~\cite{Sherman1981,Painter2006,McCulloh2003}.
A particular class of such networks that minimizes flow resistance under biologically relevant constraints 
has been studied to reveal a wealth of phenomena such as phase transitions~\cite{Banavar2000, Bohn2007}, the interdependence of flow and conduit geometry~\cite{Durand2006}, and predictions about allometric scaling relations in biology~\cite{West1997b}. When the optimization models are generalized to require resilience to damage or to consider fluctuations in the load, optimal networks reproduce the reticulate network patterns observed in biological systems~\cite{Corson2010, Katifori2010}. The optimality principles that often determine these networks also appear in non-biological context, e.g., river basins~\cite{Rinaldo1998,Sinclair1996} and are  relevant for man made systems such as gas or sewage pipe networks~\cite{Mahlke2007,DaConceicaoCunha1999}.

The overall structure of biological distribution networks is to a large extent genetically determined. However, the networks are typically composed of thousands of vessels, and genetic information cannot encode the position and diameter of each individual link~\cite{Pries2011}. Instead, development relies on local feedback mechanisms where increased flow through a vascular segment will result in improved conductivity of the vessel. For example, in plant leaves, an adaptive feedback mechanism involving the phytohormone auxin, termed ``canalization," is believed to guide morphogenesis of the venation pattern~\cite{Smith2009,Scarpella2006,Verna2015}. Beyond development, such adaptive mechanisms allow organisms to modify the network structure and respond to changing environmental cues. In slime molds, adaptation to flow of nutrients leads to efficient long-range transport~\cite{Nakagaki2000,Tero2008,Tero2010}. In animal vasculature,
both development and adaptation in the adult organism are actuated by a response to vein wall shear stress~\cite{Eichmann2005,Hu2012,Kurz2001,Scianna2013,Hacking1996}.

\begin{figure}
\includegraphics[width=\columnwidth]{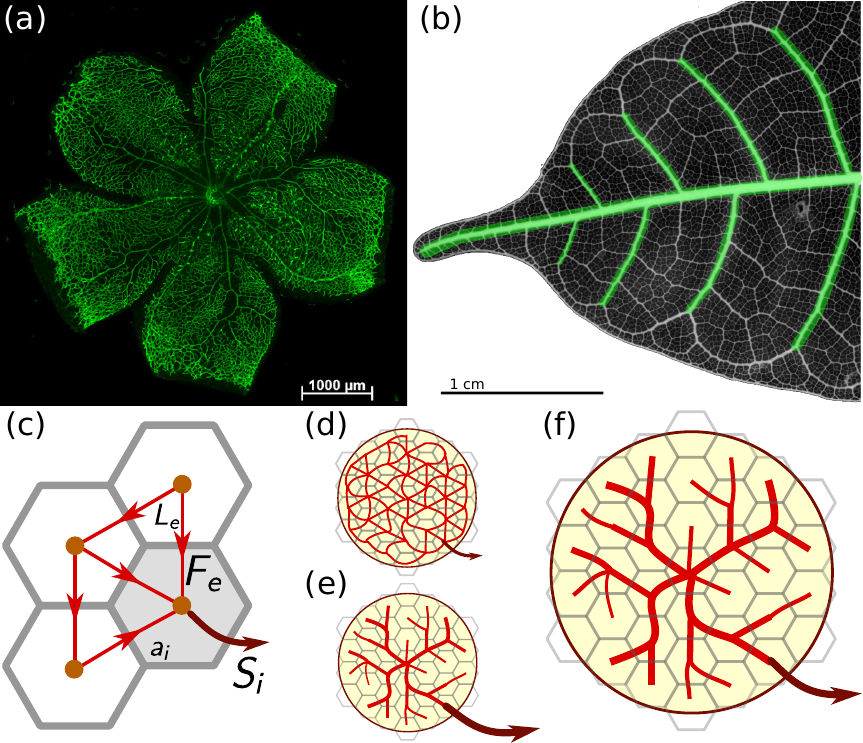}
\caption{(a) Immunostained mouse pup retina 7 days after birth. The hierarchical vascular network is clearly visible. Reprinted from~\cite{Tual-Chalot2013}, with permission. (b) Chemically cleared and stained leaf
of \emph{Protium wanningianum}. The first two levels of vascular hierarchy are marked in green.
(c) Network representation of vascular flow.
Edges (red, with lengths $L_e$) carry flows $F_e$ (arrows). 
At nodes (orange, representing areas $a_i$), net currents $S_i$ (dark red arrows, only one shown for clarity) are drawn from the network,
supplying the surrounding tissue.
(d-f) Sketch of vascular development. Overlaid in grey is a co-moving coarse-graining grid.
At the beginning (d), a tightly meshed capillary plexus is formed (capillaries in red). 
During development, the plexus is pruned and a hierarchical vascular tree is formed
either due to (e) increasing perfusion, or (f) growth.
\label{fig:fig1}}
\end{figure}
Finding the topology that optimizes flow, i.e., one that dissipates less power than others, is frequently not trivial because the objective function is often non-linear and non-convex, leading to a plethora of local optima~\cite{Bohn2007}. These local optima may perform 
significantly worse than the global optimum. 

Networks that follow simple adaptive rules have been shown to lead to steady states that are local optima of relevant objective functions ~\cite{Sinclair1996,Hu2013}. 
These states often lack structure, whereas real vasculature typically exhibits a large degree of highly symmetric hierarchical organization that can only be reproduced in models by employing global non-linear optimization techniques such as Monte-Carlo algorithms~\cite{Bohn2007} or simulated annealing~\cite{Katifori2010}. Given the evolutionary pressures for ever improved vascular networks, an important
question is how an organism is able to construct a highly optimal transport system, i.e., one that is close to the
global optimum of the relevant functional,  via local adaptive rules without recourse to direct genetic encoding of the whole pattern.

In this letter, we show that network adaptation on a growing substrate can serve as a simple, physical explanation of the globally optimized networks found in nature.
We note that the importance of growth in adaptive processes has been emphasized before in other
contexts~\cite{Lee2014,Bar-Sinai2016,Heaton2010}.
Inspired by adaptive models of vascular development in plants~\cite{Rolland-Lagan2005} and animals~\cite{Hacking1996}, we derive a set of coarse-grained dynamical equations that include scaling effects due to growth
of, e.g., the leaf blade of dicotyledonous plants or the animal embryo. Notably, the rules we derive
from growth can be interpreted as a time-dependent increase in perfusion alone (e.g., in wound healing), such that
growth may not be strictly necessary (Fig.~\ref{fig:fig1} (d-f)).
Growth manifests as local, deterministic terms in the dynamical equations, such that neither global exchange of information nor stochastic exploration of the energy landscape is necessary to produce a globally optimized pattern. In addition, the pattern is shown to be independent of initial conditions for a large part of parameter space such
that no pre-encoded pattern is necessary to guide development. Through growth, global optimization emerges from local dynamics.

In what follows, we consider a two-dimensional continuous growing sheet of tissue in which the network adapts. The results can be trivially generalized to three dimensions by modifying the scaling exponents.
This sheet may represent the leaf primordium or the surface of an organ such as the retina (Fig.~\ref{fig:fig1}).
For simplicity, growth is taken to be isotropic and uniform such that the distance between two reference
points evolves in time as $d(t) = \lambda_t d(0)$, with the time dependent scaling factor $\lambda_t$.
Correspondingly, areas scale as $A(t) = \lambda_t^2 A(0)$.
Without loss of generality, we take all dynamical quantities $x(t)$ to scale with growth as 
$x(t) = \lambda_t^\sigma x'(t)$ for some $x$-dependent exponent $\sigma$.
We model no back-reaction of the network on the growth process.

At the start of the adaptation process, we partition the sheet, e.g., with a lattice or a Voronoi tessellation.
Over the whole growth process, this partitioning will remain topologically fixed,
providing a ``co-moving frame" network that is the dual of the tessellation. 
When growth is uniform, 
the nodes of the lattice at each point in time represent a fixed
fraction (a unit) of the tissue sheet, representing 
an increasing area (Fig.~\ref{fig:fig1}).

We consider coarse-grained dynamics of quantities that flow through the network.
In animal vasculature, this is blood; during leaf morphogenesis, the
phytohormone auxin \cite{Supplement}. 
The flow $F_e$ between two tessellation units $i$ and $j$ connected
by the oriented edge $e$ can be described by $F_{e} = K_e (p_j - p_i)/L_e$,
where $p_i$ is the potential (i.e., blood pressure or morphogen concentration)
at unit $i$, $L_e$ is the distance over which the potential varies,
and $K_e$ is the conductivity.

In plants, proteins embedded in the cell membranes are responsible
for transporting auxin~\cite{Blakeslee2005,Kramer2006} with facilitated diffusion constants $K_e$.
In animals, blood flows through cylindrical vessels of radius $R_e$ according to Poiseuille's
law $K_e = z\, R_e^4$ with a constant $z$~\cite{Hacking1996,Hu2012}.

Let $\Delta: \mathcal N \rightarrow \mathcal E$ be the network's oriented incidence
matrix which maps from the node vector space $\mathcal N$ to the edge vector space $\mathcal E$.
We define the flow vector $\mathbf F \in \mathcal E$ with components $F_e$,
\begin{align}
	\mathbf F = K L^{-1}\; \Delta \mathbf p,
    \label{eq:flow}
\end{align}
where $\mathbf p = \lambda_t^\nu \mathbf p' \in \mathcal N$ is the potential
vector with components $p_i$ and $\nu$ is an (unknown) scalar.
The diagonal matrix $L = \lambda_t L'$ contains the distances, 
and the diagonal matrix $K=\lambda_t^\tau K'$ is the dynamically adapting 
conductivity the scaling $\tau$ of which will be deduced later.

The flow balance at each node reads
\begin{align}
	\Delta^T \mathbf F = \mathbf S,
    \label{eq:aux-balance}
\end{align}
where $\Delta^T$ is the transpose of the incidence matrix and $\mathbf S = \lambda_t^\delta \mathbf S'$ is a source term.
Equation~\eqref{eq:aux-balance} is Kirchhoff's current law.
In plants, the components $S_i$ describes production rate of auxin in unit $i$, which we
take to be uniform~\citep{Dimitrov2006}. Production scales with the total area,
thus $\delta = 2$.
In animals, it describes the amount of blood perfusing the tissue represented by 
unit $i$ per time.
In effectively 2D tissues such as the retina, $\delta = 2$; when a 2D vessel network services a 3D organ 
(e.g. the cortical surface arterial vasculature and the brain), $\delta = 2 + \varepsilon$, where $\varepsilon > 0$.
Combining equations~\eqref{eq:flow} and \eqref{eq:aux-balance}, we solve for the steady state
flows and obtain
\begin{align}
	\mathbf{F} = K L^{-1} \Delta (\Delta^T K L^{-1} \Delta)^\dagger \mathbf S = \lambda_t^\delta \mathbf F',
    \label{eq:flow-sln}
\end{align}
where the dagger represents the Moore-Penrose pseudoinverse.

Generalizing~\cite{Rolland-Lagan2005,Hacking1996,Hu2013}, we propose the adaptation rule
\begin{align}
	\frac{d K_{e}}{dt} = a \left(F_{e}/\hat F \right)^{2\gamma} - b K_{e} + c.
    \label{eq:adaptation}
\end{align}
This equation describes a positive feedback mechanism. If the flow $F_e$ through
an edge is large compared to $\hat F$, its conductivity increases as controlled
by the parameters $a,\gamma$. If the flow is negligible and $K_e > c/b$, conductivity
will decrease over the time scale $b^{-1}$; if $K_e < c/b$, it will increase.
Equation~\eqref{eq:adaptation} is a generalized model of auxin canalization in plants
and adaptation to vessel wall shear stress $\tau_{e} \sim F_{e}/r_{e}^3$ in animals
(see also~\cite{Supplement}).
Note that equation~\eqref{eq:adaptation} does not explicitly model tip growth,
which can be important for the growth of some networks (e.g.,~\cite{Heaton2010}).

Eqs.~\eqref{eq:flow-sln}, \eqref{eq:adaptation} can be rewritten as the equivalent system
\begin{align}
	\mathbf F' &= K' L^{-1} \Delta (\Delta^T K' L^{-1} \Delta)^\dagger \mathbf S' \label{eq:flow-rescaled}\\
    \frac{d K_e}{dt} &= a \left(\lambda_t^{\delta} F'_{e}/\hat F\right)^{2\gamma} - b K_e + c,
    \label{eq:adaptation-rescaled}
\end{align}
where all scaling factors appear explicitly.
We see that the effect of growth is to rescale the flow. Thus equivalent results
are obtained if $\mathbf S$ is time-dependent without growth.

So far the model did not require an explicit time dependence of $\lambda_t$. In what follows, we will focus on the early stages of development when growth is exponential and assume $\lambda_t=e^{\frac{r}{2}t}$, where $r$
is the area growth rate. This is a popular continuum model of growth by cell division (for details, see~\cite{Bittig2008} and~\cite{Supplement}). The dynamics is qualitatively robust as long as $\lambda_t$ increases 
(sub-)exponentially~\cite{Supplement}.

If $\delta < 0$, adaptation is suppressed and all $K_e$ tend to a uniform value.
If $\delta = 0$, growth has no effect on adaptation dynamics.
If $\delta > 0$, which is the generic case in both plants ($\delta = 2$)
and animals ($\delta = 2+\epsilon$), the flows grow without bounds as $t\rightarrow\infty$. 
In real organisms
growth eventually stops, preventing this behavior. Thus, our model is valid only
in the earlier stages. In leaves, patterning is completed long before growth slows down~\cite{Kang2004}.

\begin{figure}
\includegraphics[width=\columnwidth]{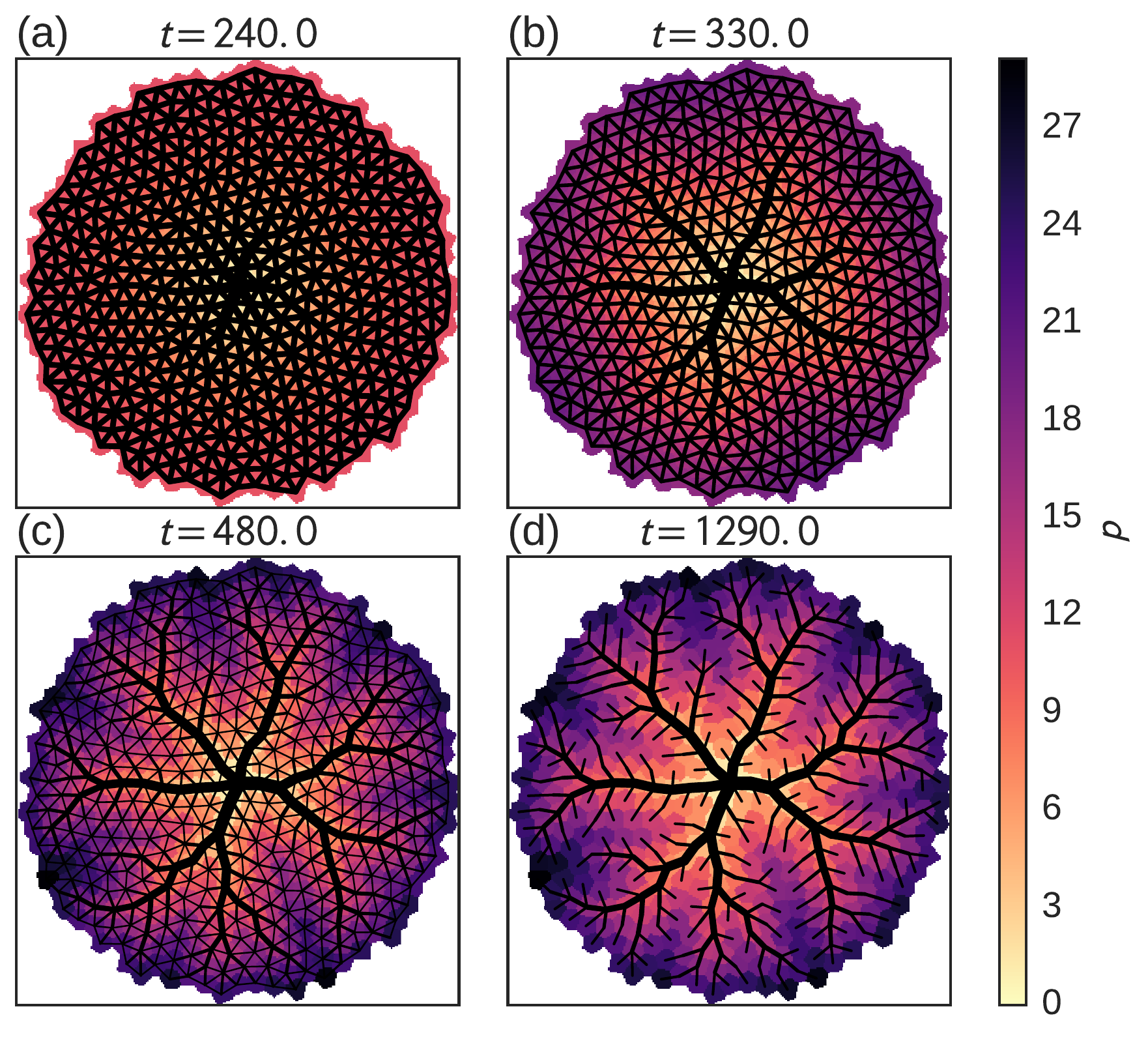}
\caption{The dynamical transients of equation~\eqref{eq:adaptation-renormalized}
for $\kappa = 1, \rho = 50$ show 
hierarchical formation of an optimized network. Line width is proportional to $(K'_e)^{1/4}$,
color is the potential $\mathbf p$, and time $t$ is measured in units of
$(b')^{-1}$. (a) The initial $K'_e$
are smoothed out and a homogeneous network is formed. (b,c) The network structure
emerges hierarchically, with largest veins first, and successively smaller veins later.
(d) After decaying of small vessels, the final network is highly organized, 
hierarchically ordered, and has a low energy.
\label{fig:transients}}
\end{figure}

We can now extract the asymptotic topology by setting
$K = \lambda_t^{2\gamma\delta} K'$. 
The dynamical equation for the asymptotic conductivities $K'$ now reads
\begin{align}
	\frac{d K'_{e}}{dt} = a \left(F'_{e}/\hat F\right)^{2\gamma} - 
    	b' K'_{e} + \lambda_t^{-2\gamma\delta}\, c,
    \label{eq:adaptation-renormalized}
\end{align}
with $b' = \left(b + \tau \dot \lambda_t/\lambda_t\right) = \left(b + r \gamma\delta\right)$.
We see that the effect of growth on the asymptotic dynamics of the topology is to
exponentially suppress the background production rate $c$ and shift the decay time scale,
where the shift by $r\gamma\delta$ comes from
the time derivative of $\lambda_t^{2\gamma\delta} K'$.
Increasing flow eventually dominates over background production.

The model is controlled by the two dimensionless parameters $\rho = b/(r\gamma\delta)$, 
the ratio between the time scales for adaptation and growth, and 
$\kappa=(c/a) (\hat F /\hat S)^{2\gamma}$, where
$c/a$ is the ratio between background growth rate and adaptation strength and the hatted quantities
are typical scales for flow and source strength.
In the rest of this Letter and all figures, we proceed to report dimensionless quantities
(see~\cite{Supplement} for details).
It can be shown (see also~\cite{Hu2013})
that for any finite $\rho$ the steady states of Eq.~\eqref{eq:adaptation-renormalized}
correspond to the critical points of the power dissipation functional
\begin{align}
	E = \sum_e L'_e \frac{(F'_e)^2}{K'_e}
    \label{eq:energy-functional}
\end{align}
under the cost constraint $\sum_e L'_e (K_e')^{\frac{1}{\gamma} - 1} \equiv \textrm{const}$.
This functional leads to realistic networks if the
constraint is concave~\cite{Bohn2007}, $1/2 < \gamma < 1$.
In the case of plants, it was shown that Eq.~\eqref{eq:energy-functional} is equivalent
to the average pressure drop~\cite{Katifori2010}, 
the physiologically relevant functional for plants~\cite{Roth-Nebelsick2001}.
In the rest of this letter, we show that for an appropriate choice of parameters,
Eq.~\eqref{eq:adaptation-renormalized}, which contains only \emph{local}, deterministic
terms, robustly leads to highly \emph{globally} optimized networks, i.e., networks
whose dissipation rate is closer to the global minimum of \eqref{eq:energy-functional}
than that of networks obtained from adaptation alone.
Thus, we demonstrate that a simple physical
mechanism such as growth can account for the remarkable optimality found in natural networks.

\begin{figure}
\includegraphics[width=\columnwidth]{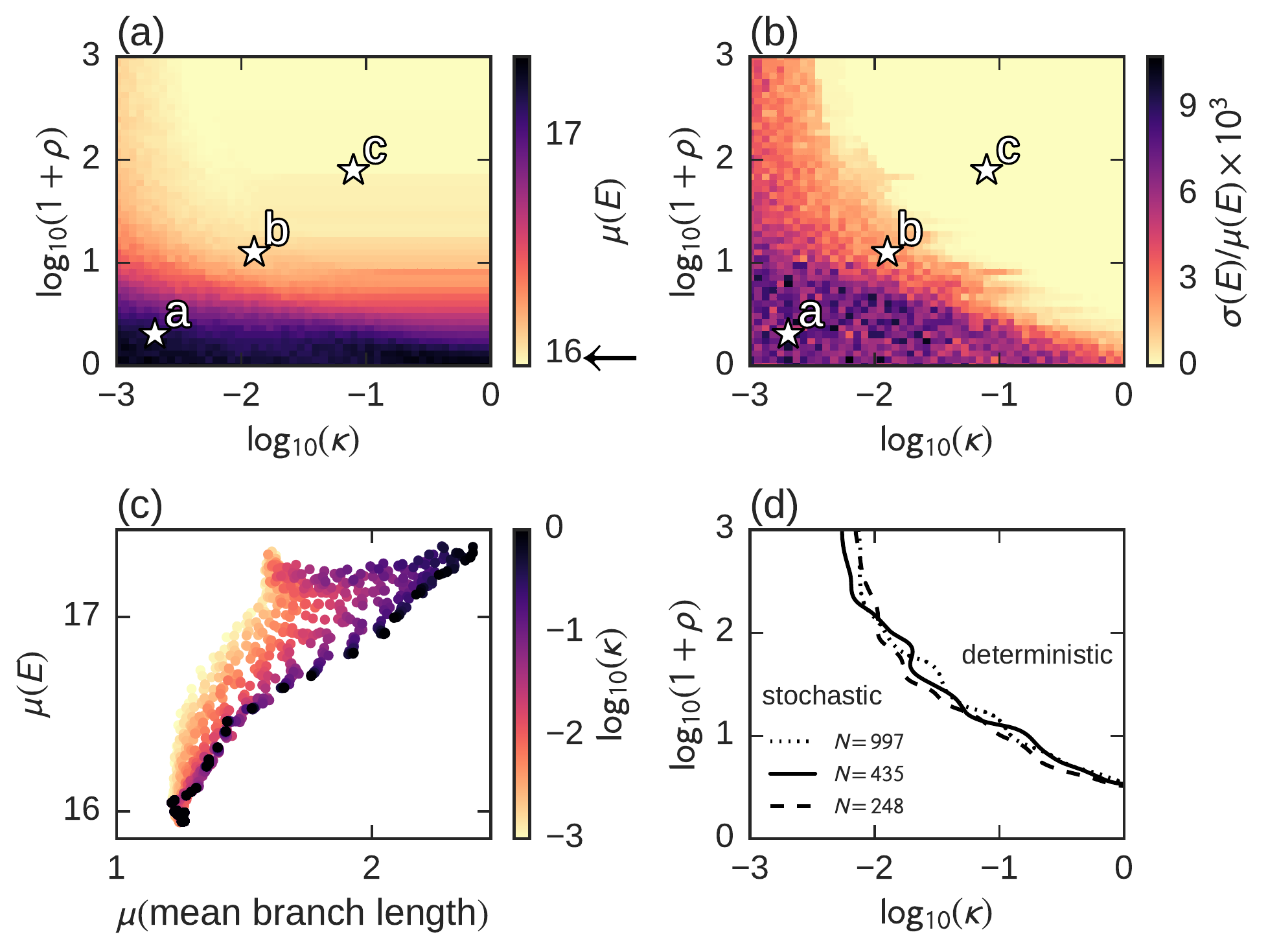}
\caption{The energy landscape shows a stochastic and a deterministic, optimized phase.
The parameter space was partitioned into a $50\times 50$ log-spaced grid and
for each parameter value, Eq.~\eqref{eq:adaptation-renormalized} was solved until convergence to
the steady state for 20
different, random initial conditions on a disordered tessellation.
(a) The mean dimensionless energy $\mu(E)$ over 20 networks. Large $\kappa$ and $\rho$ roughly correspond to
low energy, optimized networks. We mark three particular example networks that are plotted
in Fig.~\ref{fig:examples}. The minimum energy network had $E_\text{min} = 15.93$,
the the relative improvement was $(E_\text{max}-E_\text{min})/E_\text{max} = 10.3\%$.
The arrow in the colorbar marks the minimum energy solution obtained from 1000 runs of simulated annealing,
$E_\text{anneal} = 15.98$.
(b) The relative standard deviation $\sigma(E)/\mu(E)$ shows a separation of the parameter space into
a stochastic phase in which different initial conditions lead to different final networks, and a
deterministic phase in which all initial conditions are mapped to the same final state. 
We mark the same networks as in (a).
(c) Topology correlates with optimization. Highly optimized networks tend to
have a low mean branch length (mean number of edges between bifurcating nodes). The correlation
becomes stronger with increasing $\kappa$.
(d) The phase boundary in parameter space is independent of network size. 
We show the approximate position of the phase boundary by smoothing (b) and plotting the contour where
$\sigma(E)/\mu(E) = 10^{-4}$ for three tessellations.
\label{fig:phase-diagram}}
\end{figure}

\begin{figure}
\includegraphics[width=\columnwidth]{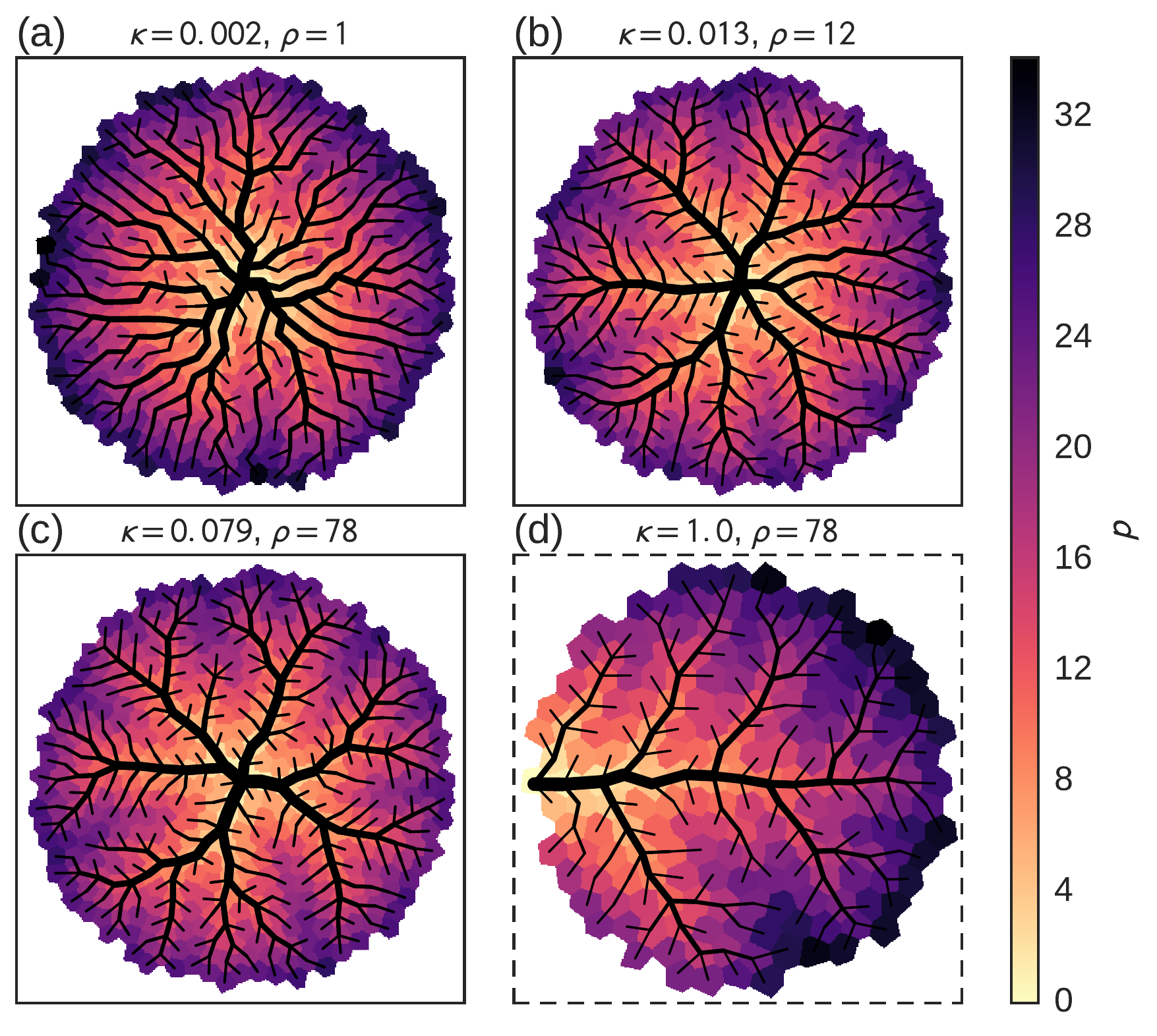}
\caption{The steady state networks marked on the energy landscape in 
Fig.~\ref{fig:phase-diagram} exhibit
a morphological transition. Line width is proportional to $(K'_e)^{1/4}$,
color is the potential $p_i$. (a) The network is disordered and not hierarchically organized, with many long branches connecting directly to the source. $E_a = 17.26$.
(b) Hierarchical organization begins to appear, $E_b = 16.19$ (c) The network is hierarchically organized 
and efficient. Few long branches are visible, $E_c = 15.95$ (compare with Fig.~\ref{fig:fig1} (a)).
(d) Optimal network when the source is at the boundary showing leaf-like
main and secondary veins (compare with Fig.~\ref{fig:fig1} (b)).
\label{fig:examples}}
\end{figure}

To mimic the randomness inherent in biological systems, we choose a disordered tessellation 
\cite{Supplement} with circular boundary and $N=435$ nodes.
The components of the source vector are
$S_i = -\delta_{0i}+\frac{1}{N-1}(1 - \delta_{0i})\frac{a_i}{\sum_{j\neq 0}a_j}$, where node $0$ is
at the center of the network and $a_i$ is the area of tessellation unit $i$ (Fig.~\ref{fig:fig1}).
We use $\gamma = 2/3$, which corresponds to both a general
model of animal vascular remodeling and a volume fixing constraint for
Eq.~\eqref{eq:energy-functional}~\cite{Supplement}.
We stress that the results do not depend on these choices, and are qualitatively
similar for other tessellations, boundary conditions, and
values of $\gamma$~\cite{Supplement}.

The network dynamics is generically characterized by two transient phases.
At first, the background production term dominates,
$\kappa\, \lambda_t^{-2\gamma\delta} \gg (F_e'/\hat S)^{2\gamma}$,
creating a homogeneous network. As production decays,
$\kappa\, \lambda_t^{-2\gamma\delta} \lesssim (F_e'/\hat S)^{2\gamma}$ holds for some
edges $e$ where flows are strongest such that the adaptive feedback takes over. 
Smaller veins are created successively as the background production is more and 
more suppressed, resulting in a hierarchically organized network (Fig.~\ref{fig:transients}
and~\cite{SupplementVideo}).
Additionally, we distinguish two phases in parameter space.
In the \emph{stochastic} phase (Fig.~\ref{fig:phase-diagram}),
the system is rapidly quenched by the adaptive feedback terms
to produce a random, non-symmetric network topology, see Fig.~\ref{fig:examples} (a).
Different initial conditions lead to different network topologies with a distribution
of energies.
In the \emph{deterministic} phase, initial smoothing persists long enough to make
the final state virtually independent of the initial conditions. Identical networks
are now obtained from different, random initial conditions in large areas of 
parameter space (Fig.~\ref{fig:phase-diagram} (b,d)).
The position of the
phase boundary is largely independent of network size (Fig.~\ref{fig:phase-diagram} (d)) unless
the type of tessellation or the boundary conditions are changed radically \cite{Supplement}.
Adaptation with growth acts as a highly efficient, deterministic optimization procedure, and can
find an energetically comparable minimum to
simulated annealing (Fig.~\ref{fig:phase-diagram} (a)).
The topology of such efficient networks is characterized by the tendency to reuse the same edge to supply
large parts of the network, as opposed to directly connecting each node to the source.
This is reflected in the the mean number of edges between two 
bifurcations (the mean branch length). Efficient networks tend to exhibit fewer non-branching nodes
(Fig.~\ref{fig:phase-diagram} (c)).
Temporally fluctuating sources  (similar to ~\cite{Hu2013}) during the adaptive process can produce loops~\cite{Supplement},
reminiscent of real reticulate biological networks. 
In addition, variable branching angles~\cite{Durand2006}, 
growth anisotropies and steric effects~\cite{Supplement} may also play a role.

We presented a dynamical model of coarse-grained network adaptation
that takes into account effects of overall network growth or, equivalently, 
increasing source strength.
We demonstrated that the parameter space of asymptotic network patterns
exhibits a stochastic and a deterministic phase.
The deterministic states were shown to often provide excellent low energy networks in the sense
of network optimization. This suggests a simple
physical mechanism such as growth may have been selected for over the course
of evolution to produce highly optimized venation in plants
and animals. 
Growth effectively reduces the dimension of the evolutionary search space
to two parameters that can be used to explore the energy landscape.
Studying appropriate mutants in plants (e.g., similar to~\cite{Sawchuk2013,Verna2015})
or animals could verify our model.
Finally, we suggest that similar biologically inspired dynamics could help improve
solutions to other global optimization problems.

\begin{acknowledgments}
This work was supported in part by the NSF award PHY-1554887 and the Burroughs Wellcome Career Award.
We wish to thank Douglas C. Daly for providing access to the leaf specimen shown in Fig.~\ref{fig:fig1} (b).
\end{acknowledgments}

\bibliographystyle{apsrev4-1}
\bibliography{references}

\end{document}